\documentclass[twocolumn,A4]{article}
\usepackage[dvips]{graphics}
\begin{document}
\onecolumn
\begin{center}
{\bf{\Large Magnetic field controlled electron transport in a thin 
cylinder}}\\
~\\
Santanu K. Maiti$^{1,2,*}$ \\
~\\
{\em $^1$Theoretical Condensed Matter Physics Division,
Saha Institute of Nuclear Physics, \\
1/AF, Bidhannagar, Kolkata-700 064, India \\
$^2$Department of Physics, Narasinha Dutt College,
129, Belilious Road, Howrah-711 101, India} \\
~\\
{\bf Abstract}
\end{center}
We explore electron transport in a thin cylinder, attached to two 
semi-infinite one-dimensional metallic electrodes, in the presence of 
both longitudinal and transverse magnetic fluxes. A simple tight-binding 
model is used to describe the system, where all the calculations are 
performed in the Green's function formalism. Quite surprisingly it is 
observed that, typical current amplitude oscillates as a function of 
the transverse magnetic flux, associated with conductance-energy 
characteristics, showing $N\phi_0$ flux-quantum periodicity, where $N$ 
and $\phi_0$ $(=ch/e)$ correspond to the system size and elementary 
flux-quantum respectively. The analysis might be helpful in fabricating 
mesoscopic switching devices, where a particular response can be delayed 
by tuning the system size $N$.
\vskip 1cm
\begin{flushleft}
\end{flushleft}
\vskip 4in
\noindent
{\bf ~$^*$Corresponding Author}: Santanu K. Maiti 

Electronic mail: santanu.maiti@saha.ac.in

\newpage
\twocolumn

\section{Introduction}

The sensitivity of electronic transport in low-dimensional systems like 
quantum dots~\cite{cron,holl}, mesoscopic loops~\cite{kulik1,kulik2}, 
quantum wires~\cite{orella1,orella2}, etc., on their geometry makes them 
truly unique in offering the possibility of electron transport in a very 
tunable way. These systems provide several anomalous characteristics in 
electron transport due to their reduced system dimensionality and quantum 
confinement. In the present age of nanoscience and technology, electronic 
devices made from quantum confined model systems are used extensively in 
fabrication of nano-electronic circuits where electron transport becomes 
predominantly coherent~\cite{nitzan1,nitzan2}. The progress of theoretical 
description in a bridge system has been followed based on the pioneering 
work of Aviram and Ratner~\cite{aviram}. Later, the actual mechanisms
underlying such transport become much more clearly resolved after several
nice experimental observations~\cite{tali,reed1,reed2} in different bridge 
systems. Though in literature many theoretical as well as experimental 
papers on electron transport are available, yet lot of controversies are 
still present between the theory and experiment, and the complete knowledge 
of the conduction mechanism in this scale is not very well established 
even today. Several key factors are there which control the electron 
transport across a bridge system, and all these effects have to 
be taken into account properly to characterize the transport. For our 
illustrative purposes, here we describe very briefly some of these factors.
(I) The quantum interference effect~\cite{baer1,baer2,baer3,tagami,walc1}
of electron waves passing through different arms of a conducting element 
which bridges two semi-infinite one-dimensional metallic electrodes, viz, 
source and drain becomes the most significant issue in electron transport 
in a bridge system. (II) The coupling of the electrodes with the bridging 
material provides an important signature in the determination of current 
amplitude~\cite{baer1}. The understanding of this coupling under 
non-equilibrium condition is a major challenge in this particular field, 
and we should take care about it in fabrication of any electronic device.    
(III) The geometry of the conducting material between the two electrodes 
itself is an important issue to control the electron transmission. To 
emphasize it, Ernzerhof {\em et al.}~\cite{ern2} have predicted several 
model calculations and provided some new significant results.
(IV) The dynamical fluctuation in the small-scale devices is another 
important factor which plays an active role and can be manifested through 
the measurement of {\em shot noise}, a direct consequence of the quantization 
of charge~\cite{blanter,walc2}. Beside these factors, several other 
parameters of the Hamiltonian that describe a system also provide 
significant effects in the determination of the current across a bridge
system.

In the present paper, we will investigate the electron transport 
properties of a thin cylinder (see Fig.~\ref{strip}), attached to two 
semi-infinite one-dimensional metallic electrodes, in the presence of 
both longitudinal and transverse magnetic fluxes, $\phi_l$ and $\phi_t$ 
respectively. A simple tight-binding model is used to illustrate the 
system, where all the calculations are performed in the Green's function
formalism. 
Our numerical study shows that, typical current amplitude across the 
cylinder oscillates as a function of the transverse magnetic flux 
$\phi_t$, associated with the radial magnetic field $B_r$, showing 
$N\phi_0$ flux-quantum periodicity instead of simple $\phi_0$-periodicity, 
where $N$ and $\phi_0$ $(=ch/e)$ correspond to the size of the cylinder and 
the elementary flux-quantum respectively. This oscillatory behavior provides 
an important signature in this particular study. It is observed that, 
for a fixed bias voltage $V$ applied by the electrodes, we can achieve 
a similar response for different cylinders by controlling the transverse 
magnetic flux associated with $B_r$. Thus, for the two cylinders of different 
sizes, similar response will be obtained for two different values of $B_r$,
depending on the system size of the cylinder. For smaller cylinder, the 
response is obtained for lower value of $B_r$, while this similar response 
will be obtained for some higher value of $B_r$ for the larger cylinder.
This clearly manifests that a particular response can be delayed by tuning
the system size $N$ quite significantly. This aspect may be utilized in
designing a tailor made switching device, and to the best of our knowledge,
this phenomenon has not been addressed earlier in the literature.

We organize the paper as follow. Following the introduction (Section $1$), 
in Section $2$, we present the model and the theoretical formulations for 
our calculations. Section $3$ discusses the significant results, and 
finally, we summarize our results in Section $4$.

\section{Model and the theoretical description}

This section follows the methodology for the calculation of the 
transmission probability ($T$), conductance ($g$) and current ($I$) in a 
small cylinder by using the Green's function technique.

\begin{figure}[ht]
{\centering \resizebox*{7cm}{5cm}{\includegraphics{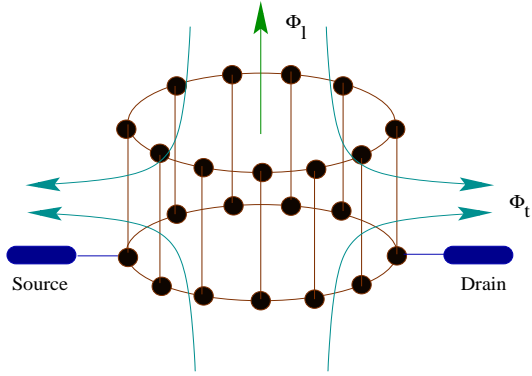}}\par}
\caption{(Color online). Schematic view of a mesoscopic cylinder attached
to two semi-infinite one-dimensional metallic electrodes and subjected
to both the longitudinal and transverse magnetic fluxes $\phi_l$ and
$\phi_t$ respectively. Filled circles correspond to the position of
the atomic sites.}
\label{strip}
\end{figure}

We begin by referring to Fig.~\ref{strip}, where a cylinder is attached 
to two semi-infinite one-dimensional metallic electrodes, viz, source 
and drain. The cylinder is subjected to a longitudinal magnetic flux 
$\phi_l$, and a transverse magnetic flux $\phi_t$. For our illustrative 
purpose here we consider the simplest possible cylinder, where only two 
isolated one-channel rings are connected by some vertical bonds. The 
transverse magnetic flux $\phi_t$ is expressed in terms of the radial 
magnetic field $B_r$ by the relation $\phi_t=B_r L d$, where the symbols 
$L$ and $d$ correspond to the circumference of the ring and the hight of 
the cylinder respectively. 

At sufficient low temperature and bias voltage, we use the Landauer
conductance formula~\cite{datta,marc} to calculate the conductance $g$
of the cylinder which can be expressed as,
\begin{equation}
g=\frac{2e^2}{h} T
\label{equ1}
\end{equation}
where $T$ becomes the transmission probability of an electron through
the cylinder. It can be expressed in terms of the Green's function of 
the cylinder and its coupling to the two electrodes by the 
relation~\cite{datta,marc},
\begin{equation}
T=Tr\left[\Gamma_S G_{c}^r \Gamma_D G_{c}^a\right]
\label{equ2}
\end{equation}
where $G_{c}^r$ and $G_{c}^a$ are respectively the retarded and advanced
Green's functions of the cylinder including the effects of the electrodes.
The parameters $\Gamma_S$ and $\Gamma_D$ describe the coupling of the
cylinder to the source and drain respectively, and they can be defined in 
terms of their self-energies. For the complete system i.e., the cylinder
with the two electrodes the Green's function is defined as,
\begin{equation}
G=\left(\epsilon-H\right)^{-1}
\label{equ3}
\end{equation}
where $\epsilon=E+i\eta$. $E$ is the injecting energy of the source electron
and $\eta$ gives an infinitesimal imaginary part to $\epsilon$. Evaluation
of this Green's function requires the inversion of an infinite matrix as the
system consists of the finite cylinder and the two semi-infinite electrodes.
However, the entire system can be partitioned into sub-matrices corresponding
to the individual sub-systems and the Green's function for the cylinder
can be effectively written as,
\begin{equation}
G_c=\left(\epsilon-H_c-\Sigma_S-\Sigma_D\right)^{-1}
\label{equ4}
\end{equation}
where $H_c$ is the Hamiltonian of the cylinder which can be written in the
tight-binding model within the non-interacting picture like,
\begin{eqnarray}
H_c & = & \sum_{i=1}^N \epsilon_i^L c_i^{L\dagger} c_i^L  
+ \sum_{i=1}^N \epsilon_i^U c_i^{U\dagger} c_i^U \nonumber \\
  & + & v_l^L \sum_{<ij>}
\left[e^{i\left( \theta_1 - \theta_2 \right)} c_i^{L\dagger} c_j^L
+ e^{-i \left( \theta_1 - \theta_2 \right)} c_j^{L\dagger} 
c_i^L \right] \nonumber \\
  & + & v_l^{U} \sum_{<ij>}
\left[e^{i\left( \theta_1 + \theta_2 \right)} c_i^{U\dagger} c_j^U
+ e^{-i \left( \theta_1 + \theta_2 \right)} c_j^{U\dagger} 
c_i^U \right] \nonumber \\
 & + & v_t \sum_{i=1}^N \left(c_i^{L\dagger}c_i^U + c_i^{U\dagger}
c_i^L \right)
\label{equ5}
\end{eqnarray}
In the above Hamiltonian ($H_c$), $\epsilon_i^L$'s ($\epsilon_i^U$'s) are
the site energies in the lower (upper) ring, $c_i^{L\dagger}$
($c_i^{U\dagger}$) is the creation operator of an electron at site $i$
in the lower (upper) ring, and the corresponding annihilation operator for
this site $i$ is denoted by $c_i^L$ ($c_i^U$). The symbol $v_l^L$ ($v_l^U$)
gives the nearest-neighbor hopping integral in the lower (upper) ring,
while the parameter $v_t$ corresponds to the transverse hopping strength
between the two rings of the cylinder. $\theta_1$ and $\theta_2$ are the
two phase factors those are related to the longitudinal and transverse
fluxes by the expressions, $\theta_1=2 \pi \phi_l/N \phi_0$ and
$\theta_2=\pi \phi_t/N \phi_0$, where $N$ represents the total number of
atomic sites in each ring and $\phi_0=ch/e$ is the elementary flux-quantum. 
Similar kind of tight-binding Hamiltonian is also used to describe the 
two semi-infinite
one-dimensional perfect electrodes where the Hamiltonian is parametrized
by constant on-site potential $\epsilon_0$ and nearest-neighbor hopping
integral $t_0$. In Eq.~\ref{equ4}, $\Sigma_S=h_{Sc}^{\dagger}g_S h_{Sc}$
and $\Sigma_D=h_{Dc} g_D h_{Dc}^{\dagger}$ are the self-energy operators
due to the two electrodes, where $g_S$ and $g_D$ correspond to the Green's
functions of the source and drain respectively. $h_{Sc}$ and $h_{Dc}$ are
the coupling matrices and they will be non-zero only for the adjacent
points of the cylinder, and the electrodes respectively. The matrices 
$\Gamma_S$ and $\Gamma_D$ can be calculated through the expression,
\begin{equation}
\Gamma_{S(D)}=i\left[\Sigma_{S(D)}^r-\Sigma_{S(D)}^a\right]
\label{equ6}
\end{equation}
where $\Sigma_{S(D)}^r$ and $\Sigma_{S(D)}^a$ are the retarded and advanced 
self-energies respectively, and they are conjugate with each other.
Datta {\em et. al.}~\cite{tian} have shown that the self-energies can be
expressed like as,
\begin{equation}
\Sigma_{S(D)}^r=\Lambda_{S(D)}-i \Delta_{S(D)}
\label{equ7}
\end{equation}
where $\Lambda_{S(D)}$ are the real parts of the self-energies which
correspond to the shift of the energy eigenvalues of the cylinder and
the imaginary parts $\Delta_{S(D)}$ of the self-energies represent the
broadening of these energy levels. This broadening is much larger than the
thermal broadening and this is why we restrict our all calculations only
at absolute zero temperature. All the informations about the
cylinder-to-electrode coupling are included into these two self-energies.

The current passing across the cylinder is depicted as a single-electron
scattering process between the two reservoirs of charge carriers. The
current $I$ can be computed as a function of the applied bias voltage $V$
through the relation~\cite{datta},
\begin{equation}
I(V)=\frac{e}{\pi \hbar}\int_{E_F-eV/2}^{E_F+eV/2} T(E,V) dE
\label{equ8}
\end{equation}
where $E_F$ is the equilibrium Fermi energy. For the sake of simplicity,
we assume that the entire voltage is dropped across the cylinder-electrode
interfaces and this assumption doesn't greatly affect the qualitative aspects
of the $I$-$V$ characteristics. Such an assumption is based on the fact that,
the electric field inside the cylinder especially for small cylinders seems
to have a minimal effect on the conductance-voltage characteristics. On the
other hand, for quite larger cylinders and high bias voltages the electric
field inside the cylinder may play a more significant role depending on the
internal structure and size of the cylinder~\cite{tian}, yet the effect
is quite small.

\section{Results and discussion}

To reveal the basic mechanisms of $\phi_l$ and $\phi_t$ on the electron 
transport, here we present all the results only for the non-interacting 
electron picture. With this assumption, the model becomes quite simple 
and all the basic features can be well understood. 
Another realistic assumption is that, we focus on the perfect cylinders 
only i.e., the site energies are taken as $\epsilon_i^L=\epsilon_i^U=0$ 
for all $i$. As illustrative purposes, we parametrize all the hopping 
integrals ($v_l^L$, $v_l^U$ and $v_t$) of the cylinder by a single 
parameter $v$, viz, we set $v_l^L=v_l^U=v_t=v$. In the present article, 
we concentrate our study of electron transport in the two separate 
\begin{figure}[ht]
{\centering \resizebox*{7cm}{8.5cm}{\includegraphics{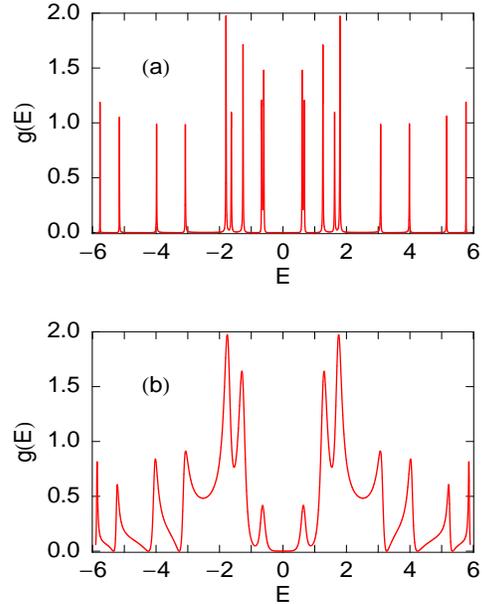}}\par}
\caption{(Color online). $g$-$E$ characteristics for a thin cylinder with
$N=12$, $\phi_l=0.2$ and $\phi_t=2$. (a) weak-coupling limit and
(b) strong-coupling limit.}
\label{cond}
\end{figure}
regimes. These are so-called the weak-coupling and strong-coupling 
regimes respectively. The weak-coupling regime is specified by the 
condition $\tau_{\{S,D\}} << v$. On the other hand, the condition 
$\tau_{\{S,D\}}\sim v$ is used to denote the strong-coupling regime. 
Here, the parameters $\tau_S$ and $\tau_D$ correspond to the hopping 
strengths of the cylinder to the source and drain respectively. In these 
two limiting cases, we choose the values of the different parameters as 
follow: $\tau_S=\tau_D=0.5$, $v=2.5$ (weak-coupling) and $\tau_S=\tau_D=2$, 
$v=2.5$ (strong-coupling). The on-site energy $\epsilon_0$ is set to $0$ 
(we can take any constant value of it instead of zero, since it provides 
only the reference energy level) for the electrodes, and the hopping 
strength $t_0$ is taken as $3$ in the two semi-infinite metallic 
electrodes. For the sake of simplicity, we set the Fermi energy $E_F=0$ 
and choose the units where $c=e=h=1$.

Let us begin our discussion with the variation of the conductance $g$ as
a function of the injecting electron energy $E$. As illustrative 
examples, in Fig.~\ref{cond} we plot the conductance-energy ($g$-$E$)
characteristics for a thin cylinder with $N=12$, considering the 
parameters $\phi_l=0.2$ and $\phi_t=2$. Figures~\ref{cond}(a) and (b)
correspond to the results for the weak- and strong-coupling regimes
respectively. It shows that, in the limit of weak-coupling, the 
conductance exhibits fine resonance peaks for some particular
energies, while for all other energies the conductance almost disappears. 
Out of these resonance peaks, only for the two energies the conductance
approaches the value $2$. Thus, for these energies the transmission 
\begin{figure}[ht]
{\centering \resizebox*{7cm}{4.5cm}{\includegraphics{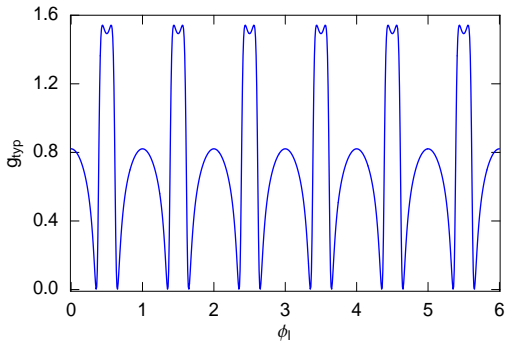}}\par}
\caption{(Color online). Typical conductance as a function of $\phi_l$ for
a thin cylinder with $N=12$ in the strong-coupling limit, where the
parameter $\phi_t$ is set to $2$. The conductance is calculated at the
energy $E=2.1$.}
\label{condlong}
\end{figure}
probability $T$ becomes unity, since we get the relation $g=2T$ from 
the Landauer conductance formula (see Eq.~\ref{equ1} with $e=h=1$). 
On the other hand, for other resonances, the conductance does not reach 
to $2$ any more, and gets much reduced value. This is due to the quantum 
interference effect of the electronic waves passing through the different 
arms of the cylinder, and it can be interpreted as follow. During the 
motion of the electrons from the source to drain through the cylinder, 
the electron waves propagating along the different possible pathways can 
get a phase shift among themselves, according to the result of quantum 
interference.
Therefore, the probability amplitude of getting an electron across the
cylinder either becomes strengthened or weakened. This causes the
transmittance cancellations and provides anti-resonances in the
conductance spectrum. Thus it can be emphasized that the electron
transmission is strongly affected by the quantum interference effects,
and hence the cylinder to electrodes interface structure. Now all these
resonance peaks are associated with the energy eigenvalues of the
cylinder, and accordingly, we can say that the conductance spectrum 
manifests itself the electronic structure of the cylinder. In the
resonance spectrum we get more peaks in the presence of both $\phi_l$
and $\phi_t$ compared to those as obtained when both these two fluxes
are identically zero. This is due to the fact that the fluxes $\phi_l$
\begin{figure}[ht]
{\centering \resizebox*{7cm}{10cm}{\includegraphics{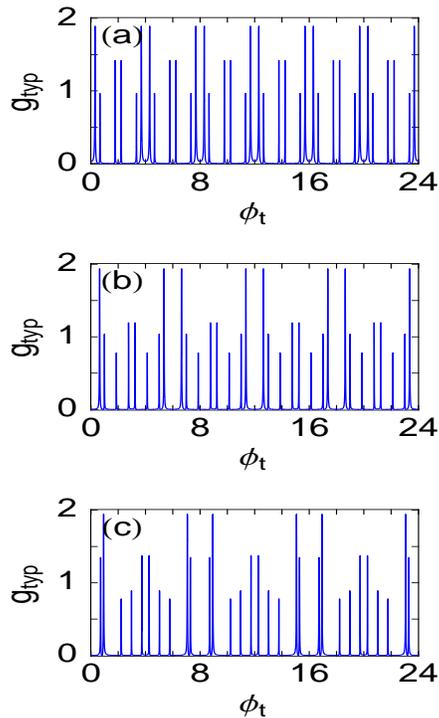}}\par}
\caption{(Color online). Typical conductances as a function of $\phi_t$
for three different cylinders with (a) $N=4$, (b) $N=6$ and (c) $N=8$
respectively, in the limit of weak-coupling. The energy $E$ is fixed
at $2.1$ and the parameter $\phi_l$ is set to $0.2$.}
\label{condtran}
\end{figure}
and $\phi_t$ remove all the degeneracies in the energy levels which 
provide more resonance peaks in the conductance spectrum. With these
features, we get additional one feature when the coupling strength
of the cylinder to the electrodes increases from the low regime to
high one. In the limit of strong cylinder-to-electrode coupling, all 
these resonances get substantial widths compared to the weak-coupling 
limit. The results are shown in Fig.~\ref{cond}(b). The contribution for 
the broadening of the resonance peaks in this strong-coupling limit
appears from the imaginary parts of the self-energies $\Sigma_S$ and
$\Sigma_D$ respectively~\cite{datta}, as mentioned earlier. Thus by 
tuning the coupling strength, we can get the electron transmission 
across the cylinder for wider range of energies and it provides an 
important signature in the study of current-voltage characteristics. 

In order to express the dependence of both the two different fluxes 
on the electron transport much more clearly,
in Figs.~\ref{condlong} and \ref{condtran} we display the variation of 
the typical conductances as a function of  $\phi_l$ and $\phi_t$ 
respectively. Figure~\ref{condlong} shows the variation of the typical 
conductance ($g_{typ}$) with the longitudinal magnetic flux $\phi_l$ in 
the limit of strong-coupling for a cylinder considering $N=12$. Here we 
\begin{figure}[ht]
{\centering \resizebox*{7cm}{8.5cm}{\includegraphics{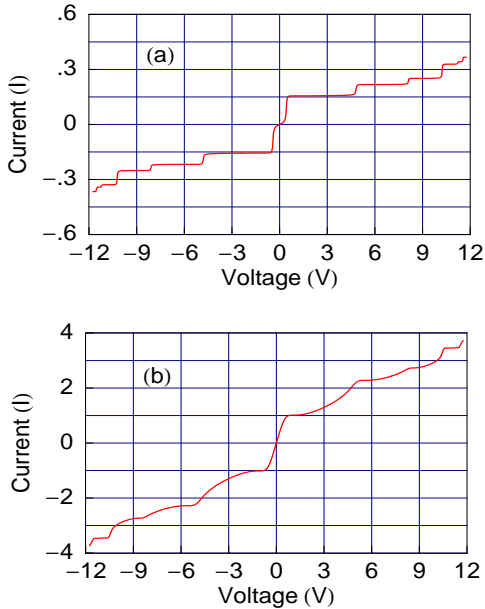}}\par}
\caption{(Color online). $I$-$V$ characteristics for a thin cylinder
with $N=6$, $\phi_l=0.2$ and $\phi_t=2$. (a) weak-coupling limit and
(b) strong-coupling limit.}
\label{current}
\end{figure}
calculate $g_{typ}$ for the energy $E=2.1$, when the parameter $\phi_t$ 
is set to $2$. It shows that the typical conductance exhibits oscillatory
behavior with $\phi_l$ showing $\phi_0$ ($=1$, since $c=e=h=1$ in our 
present formulation) flux-quantum periodicity. This $\phi_0$ periodicity 
is well known in the 
literature. The remarkable behavior is observed only for the case when 
we apply the transverse magnetic flux in the cylinder. For the illustrative 
purposes, in Fig.~\ref{condtran} we plot the variation of the typical 
conductance of the three different cylinders in the limit of weak-coupling, 
where (a), (b) and (c) correspond to the results for the cylinders with 
$N=4$, $6$ and $8$ respectively. In this case, we set the energy $E=2.1$ 
and fix $\phi_l$ to $0.2$. Quite interestingly it is observed that, the 
typical conductance varies periodically with the flux $\phi_t$ showing 
$N\phi_0$ flux-quantum periodicity, instead of the traditional $\phi_0$ 
flux-quantum periodicity as observed in the previous case. The detailed
investigation reveals that the nature of the $g_{typ}$ vs. $\phi_t$
characteristic within a single period of any one of these three cylinders
is quite similar to that of the other two cylinders within their
respective periods. Thus the positions of the resonance peaks shift 
appropriately with the increase of the system size $N$ to make the 
nature invariant. This is really a very interesting phenomenon and can 
be utilized to get a delayed response just by tuning the system size $N$.

All these features of electron transport become much more clearly visible
by studying the current-voltage ($I$-$V$) characteristics. The current 
\begin{figure}[ht]
{\centering \resizebox*{7cm}{4.5cm}{\includegraphics{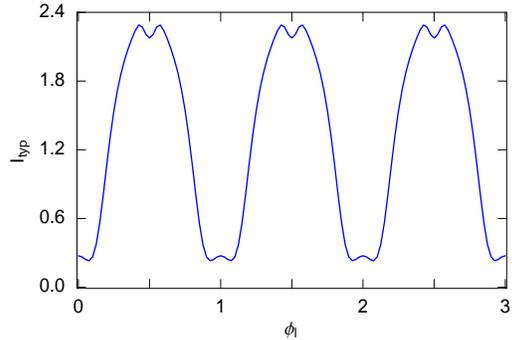}}\par}
\caption{(Color online). Typical current amplitude as a function of $\phi_l$
for a thin cylinder with $N=6$ in the strong-coupling limit, where the
parameter $\phi_t$ is set to $2$. The current is calculated at the bias
voltage $V=1.95$.}
\label{currlong}
\end{figure}
across the cylinder is computed from the integration procedure of the 
transmission function $T$ as prescribed in Eq.~\ref{equ8}. The transmission 
function varies exactly similar to that of the conductance spectrum, differ 
only in magnitude by the factor $2$ since the relation $g=2T$ holds from 
the Landauer conductance formula Eq.~\ref{equ1}. As representative examples,
in Fig.~\ref{current} we plot the current-voltage characteristics for a
thin cylinder with $N=6$, $\phi_l=0.2$ and $\phi_t=2$, where (a) and (b)
represent the results for the weak and strong cylinder-to-electrode coupling
limits respectively. In the limit of weak-coupling, the current exhibits
staircase-like structure with fine steps as a function of the applied bias 
voltage. This is due to the existence of the sharp resonance peaks in the 
conductance spectrum in this limit of coupling, since the current is 
computed by the integration method of the transmission function $T$. With 
the increase of the applied bias voltage, the electrochemical potentials 
on the electrodes are shifted gradually, and finally cross one of the 
quantized energy levels of the cylinder. Accordingly, a current channel 
is opened up and the current-voltage characteristic curve provides a jump. 
On the other hand, for the strong cylinder-to-electrode coupling, the 
current varies almost continuously with the applied bias voltage and 
achieves much larger amplitude than the weak-coupling case. The reason
is that, in the limit of strong-coupling all the energy levels get
\begin{figure}[ht]
{\centering \resizebox*{7cm}{8.5cm}{\includegraphics{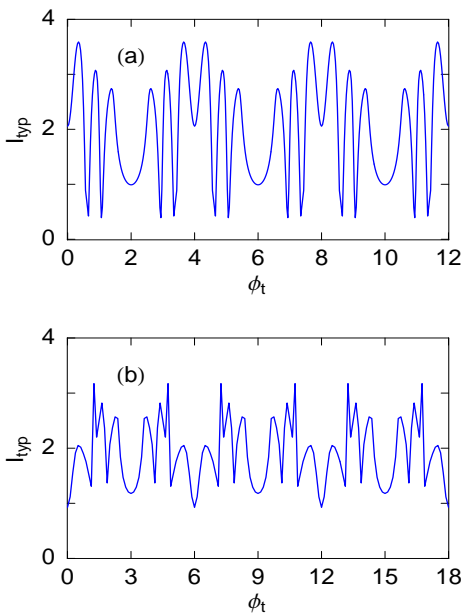}}\par}
\caption{(Color online). Typical current amplitudes as a function of
$\phi_t$ for two different cylinders with (a) $N=4$ and (b) $N=6$
respectively, in the limit of strong-coupling. The voltage $V$ is
fixed at $0.7$ and the parameter $\phi_l$ is set to $0.3$.}
\label{currtran}
\end{figure}
broadened which provide larger current in the integration procedure
of the transmission function $T$. Thus by tuning the strength of the 
cylinder-to-electrode coupling, we can achieve very large current 
amplitude from the very low one for the same bias voltage $V$.

Finally, we concentrate our study on the variation of the typical current 
amplitude ($I_{typ}$) with the magnetic fluxes $\phi_l$ and $\phi_t$ to
illuminate the significant effects of these fluxes on the electron transport. 
In Fig.~\ref{currlong}, we display the variation of $I_{typ}$ as a function 
of $\phi_l$
for a thin cylinder with $N=6$ in the limit of strong-coupling, where
the parameter $\phi_t$ is fixed to $2$. The typical current is computed
at the bias voltage $V=1.95$. It predicts that the current amplitude shows 
an oscillatory behavior with $\phi_l$, giving $\phi_0$ flux-quantum 
periodicity. This is completely analogous to the variation of $g_{typ}$
with $\phi_l$ as described earlier. The interesting feature is observed
only when we study the variation of $I_{typ}$ with $\phi_t$, instead of 
$\phi_l$. The results are shown in Fig.~\ref{currtran}, where (a) and
(b) correspond to the variation of the current amplitudes for the cylinders 
with $N=4$ and $6$ respectively. The typical current amplitudes are 
calculated in the limit of strong-coupling for the bias voltage $V=0.7$, 
when the parameter $\phi_l$ is set to $0.3$. Quite interestingly we see 
that, the current amplitude varies periodically with $\phi_t$ providing 
$N\phi_0$ flux-quantum periodicity, instead of the conventional $\phi_0$
periodicity. Thus for the cylinder with $N=4$, the current shows
$4\phi_0$ ($=4$) periodicity, while it becomes $6\phi_0$ ($=6$) for the 
cylinder with $N=6$. From these results we clearly observe that the variation
of $I_{typ}$ with $\phi_t$ within a period for the cylinder with $N=6$ is
quite similar to that of the cylinder with $N=4$ within its single
period. This is just the replica of the $g_{typ}$ versus $\phi_t$ 
characteristics which we have described earlier. Thus for a fixed bias 
voltage, we can achieve the similar response for different values of $N$ 
just by tuning the flux $\phi_t$ associated with the radial magnetic 
field $B_r$. This phenomenon can be utilized as follow. Let us consider
two such cylinders of different sizes. For these two cylinders, a particular 
response is obtained for the two different values of $B_r$, depending 
on the size of these cylinders. Therefore, for the cylinder of smaller size 
the similar response is achieved at the lower value of $B_r$ compared to 
the cylinder of larger size. This clearly manifests that a particular 
response can be delayed by tuning the size $N$ of the cylinder quite
significantly. This aspect is really a very interesting, and provides 
a signature for manufacturing mesoscopic switching devices.

\section{Concluding remarks}

In conclusion, we have addressed the electron transport in a thin 
cylinder, attached to two semi-infinite one-dimensional metallic 
electrodes, in the presence of both longitudinal and transverse 
magnetic fluxes. We have used a simple tight-binding model to 
describe the system where all the calculations have been done in 
the Green's function formalism. Quite interestingly we have 
observed that the typical current amplitude oscillates as a function 
of the transverse magnetic flux $\phi_t$, associated with the 
conductance-energy characteristics, providing $N\phi_0$ flux-quantum 
periodicity. This feature is completely different from the traditional
oscillatory behavior like as we have got for the case of $I_{typ}$ versus
$\phi_l$ characteristic. Our results have predicted that a particular
response can be delayed by tuning the size of the cylinder $N$.
This aspect may be utilized in designing a tailor made mesoscopic 
switching device.

This is our first step to describe how the electron transport in a thin
cylinder can be controlled very nicely by means of the longitudinal and 
transverse magnetic fluxes. We have made several realistic assumptions 
by ignoring the effects of the electron-electron correlation, disorder, 
temperature, finite width of the electrodes, etc. Here we discuss very 
briefly about these approximations. The inclusion of the electron-electron 
correlation in the present model is a major challenge to us, since over 
the last few years many people have studied a lot to incorporate this 
effect, but no such proper theory has yet been developed. In this work, 
we have presented all the results only for the ordered systems. But in 
real samples, the presence of impurities will affect the electronic 
structure and hence the transport properties. The effect of the
temperature has already been pointed out earlier, and, it has been
examined that the presented results will not change significantly even
at finite temperature, since the broadening of the energy levels of the
cylinder due to its coupling with the electrodes will be much larger than
that of the thermal broadening~\cite{datta}. The other important assumption
is that here we have chosen the linear chains instead of wider leads,
since we are mainly interested about the basic physics of the cylinder.
Though the results presented here change with the increase of the
thickness of the leads, but all the basic features remain quite invariant.
Finally, we would like to say that we need further study in this system 
by incorporating all these effects.

\vskip 0.3in
\noindent
{\bf\Large Acknowledgments}
\vskip 0.2in
\noindent
I acknowledge with deep sense of gratitude the illuminating comments and
suggestions I have received from Prof. Arunava Chakrabarti and Prof.
S. N. karmakar during the calculations.

\end{document}